\documentclass[aps,prl,twocolumn,showpacs,superscriptaddress]{revtex4}
\usepackage{graphicx}
\usepackage{bm}

\begin{document}

\title{Spin excitation anisotropy in optimal-isovalent-doped superconductor BaFe$_2$(As$_{0.7}$P$_{0.3}$)$_2$}

\author{Ding Hu}
\email{dinghuphys@gmail.com}
\affiliation{Department of Physics and Astronomy, Rice University, Houston, Texas 77005-1827, USA}
\affiliation{Center for Advanced Quantum Studies and Department of Physics, Beijing Normal University, Beijing 100875, China}
\affiliation{Beijing National Laboratory for Condensed Matter Physics, Institute of Physics, Chinese Academy of Sciences, Beijing 100190, China}
\author{Wenliang Zhang}
\affiliation{Beijing National Laboratory for Condensed Matter Physics, Institute of Physics, Chinese Academy of Sciences, Beijing 100190, China}
\affiliation{School of Physical Sciences, University of Chinese Academy of Sciences, Beijing 100190, China}
\author{Yuan Wei}
\affiliation{Beijing National Laboratory for Condensed Matter Physics, Institute of Physics, Chinese Academy of Sciences, Beijing 100190, China}
\affiliation{School of Physical Sciences, University of Chinese Academy of Sciences, Beijing 100190, China}
\author{Bertrand Roessli}
\affiliation{Laboratory for Neutron Scattering and Imaging, Paul Scherrer Institut, CH-5232 Villigen, Switzerland}
\author{ Markos Skoulatos}
\affiliation{Laboratory for Neutron Scattering and Imaging, Paul Scherrer Institut, CH-5232 Villigen, Switzerland}
\affiliation{Heinz Maier-Leibnitz Zentrum (MLZ) and Physics Department E21,
 Technische Universit$\ddot{a}$t M$\ddot{u}$nchen, D-85748 Garching, Germany
}
\author{Louis Pierre Regnault}
\affiliation{Institut Laue Langevin, 71 Avenue des Martyrs, 38042 Grenoble, France}
\author{Genfu Chen}
\affiliation{Beijing National Laboratory for Condensed Matter Physics, Institute of Physics, Chinese Academy of Sciences, Beijing 100190, China}
\affiliation{School of Physical Sciences, University of Chinese Academy of Sciences, Beijing 100190, China}
\affiliation{Collaborative Innovation Center of Quantum Matter, Beijing, China}
\author{Yu Song}
\affiliation{Department of Physics and Astronomy, Rice University, Houston, Texas 77005-1827, USA}
\author{Huiqian Luo}
\affiliation{Beijing National Laboratory for Condensed Matter Physics, Institute of Physics, Chinese Academy of Sciences, Beijing 100190, China}
\author{Shiliang Li}
\affiliation{Beijing National Laboratory for Condensed Matter Physics, Institute of Physics, Chinese Academy of Sciences, Beijing 100190, China}
\affiliation{School of Physical Sciences, University of Chinese Academy of Sciences, Beijing 100190, China}
\affiliation{Collaborative Innovation Center of Quantum Matter, Beijing, China}
\author{Pengcheng Dai}
\email{pdai@rice.edu}
\affiliation{Department of Physics and Astronomy, Rice University, Houston, Texas 77005-1827, USA}
\affiliation{Center for Advanced Quantum Studies and Department of Physics, Beijing Normal University, Beijing 100875, China}

\begin{abstract}
We use neutron polarization analysis to study spin excitation anisotropy in the optimal-isovalent-doped superconductor BaFe$_2$(As$_{0.7}$P$_{0.3}$)$_2$ ($T_{c} = 30$ K).
Different from optimally hole and electron-doped BaFe$_2$As$_2$, where there is a clear spin excitation anisotropy in the paramagnetic tetragonal state well above $T_{c}$,
we find no spin excitation anisotropy for energies above 2 meV in the normal state of BaFe$_2$(As$_{0.7}$P$_{0.3}$)$_2$. Upon entering the superconducting state, significant spin excitation anisotropy develops at the antiferromagnetic (AF) zone center ${\bf Q}_{\rm AF}=(1,0,L=\textrm{odd})$, while magnetic spectrum is isotropy at the zone boundary ${\bf Q}=(1,0,L=\textrm{even})$.
By comparing temperature, wave vector, and polarization dependence of the
spin excitation anisotropy in BaFe$_2$(As$_{0.7}$P$_{0.3}$)$_2$
and hole-doped Ba$_{0.67}$K$_{0.33}$Fe$_2$As$_2$ ($T_{c}=38$ K), we conclude that
 such anisotropy arises from spin-orbit coupling and is associated with the nearby AF order and superconductivity.
\end{abstract}


\pacs{74.70.Xa, 74.70.-b, 78.70.Nx}

\maketitle

The spin-orbit coupling (SOC) is an interaction of an electron's spin with its motion. While the
importance of SOC to electronic properties of the $4d$ and $5d$ correlated electron
materials such as Sr$_2$RuO$_4$ and Sr$_2$IrO$_4$ is long recognized \cite{Haverkort,BJKim}, its relevance to the physics of the $3d$ correlated
electron materials such as iron pnictide superconductors is much less clear. Since
iron pnictide superconductors are derived from metallic parent compounds exhibiting antiferromagnetic (AF) order
at $T_N$  below a tetragonal-to-orthorhombic structural transition
temperature $T_s$ associated with orbital ordering and nematic phase [Fig. 1(a)] \cite{kamihara,cruz,johnston,dai,Fernandes14N}, most microscopic theories for
iron based superconductors are focused on the role of spin- \cite{hirschfeld,Scalapino,Qimiao16NRM}, orbital- \cite{Kontani}, or nematic \cite{Fernandes14N,HHKuo2016} fluctuations
to the electron pairing and superconductivity.  Although angle resolved photoemission spectroscopy (ARPES) experiments on different families of iron-based
superconductors have identified the presence of SOC through observation of electronic band splitting at the Brillouin zone center (ZC) below $T_s$ \cite{Borisenko,Johnson,Watson},
much is unknown concerning the role of SOC to the AF order, nematic phase, electron pairing mechanism and superconductivity \cite{kseo17,ovafek17,MChristensen,Fernandes14}.

In addition to its impact on the Fermi surface and electronic band dispersions, SOC also brings
lattice anisotropies into anisotropies of magnetic fluctuations 
\cite{Korshunov2013,Applegate}, as seen from nuclear magnetic resonance \cite{ZLi11} and polarized inelastic neutron scattering (INS)
experiments on different iron-based superconductors \cite{Lipscombe,prokes,PSteffens,HQLuo2013,CZhang2014,NQureshi2012,CWangPRX,CZhang2013,NQureshi2014,YSong16,Waber16,MMa17}.
Compared with ARPES measurements, polarized INS measurements have much better energy and momentum resolution, and can directly
probe the energy, wave vector, and temperature dependence of the spin excitation anisotropy and determine its relationship with $T_c$, $T_N$, $T_s$, and nematic phase.
For hole-doped Ba$_{1-x}$K$_x$Fe$_2$As$_2$ \cite{Rotter,Avci,Wasser,Allred}, electron-doped BaFe$_{2-x}TM_x$As$_2$
($TM=$ Co, Ni) \cite{ASefat,LJLi09,DKPratt2011,XYLuPRL}, and isovalent-doped BaFe$_2$(As$_{1-x}$P$_{x}$)$_2$ \cite{SJiang09,Shibauchi14,DHu2015} superconductors,
unpolarized INS experiments found that superconductivity is coupled with the appearance of a low-energy collective spin excitation mode
termed spin resonance that has superconducting order parameter-like temperature dependence below $T_{c}$ \cite{Christianson2008,MDLumsden2009,SChi2009,MWang13,CHLee13,HDing16}.  Although polarized INS experiments have conclusively established
the presence of SOC induced low-energy spin excitation anisotropy near the resonance mode
in different families of iron-based superconductors \cite{Lipscombe,prokes,PSteffens,HQLuo2013,CZhang2014,NQureshi2012,CWangPRX,CZhang2013,NQureshi2014,YSong16,Waber16,MMa17}, the
spin excitation anisotropy persists to temperatures well above $T_{N}$ and $T_{s}$ in the paramagnetic tetragonal state, and becomes
isotropic near the nematic ordering temperature \cite{PSteffens,HQLuo2013,CZhang2013,NQureshi2014,YSong16}.
Therefore, it is still unclear how SOC is coupled to spin fluctuation anisotropy and superconductivity.

\begin{figure}
\includegraphics[scale=.4]{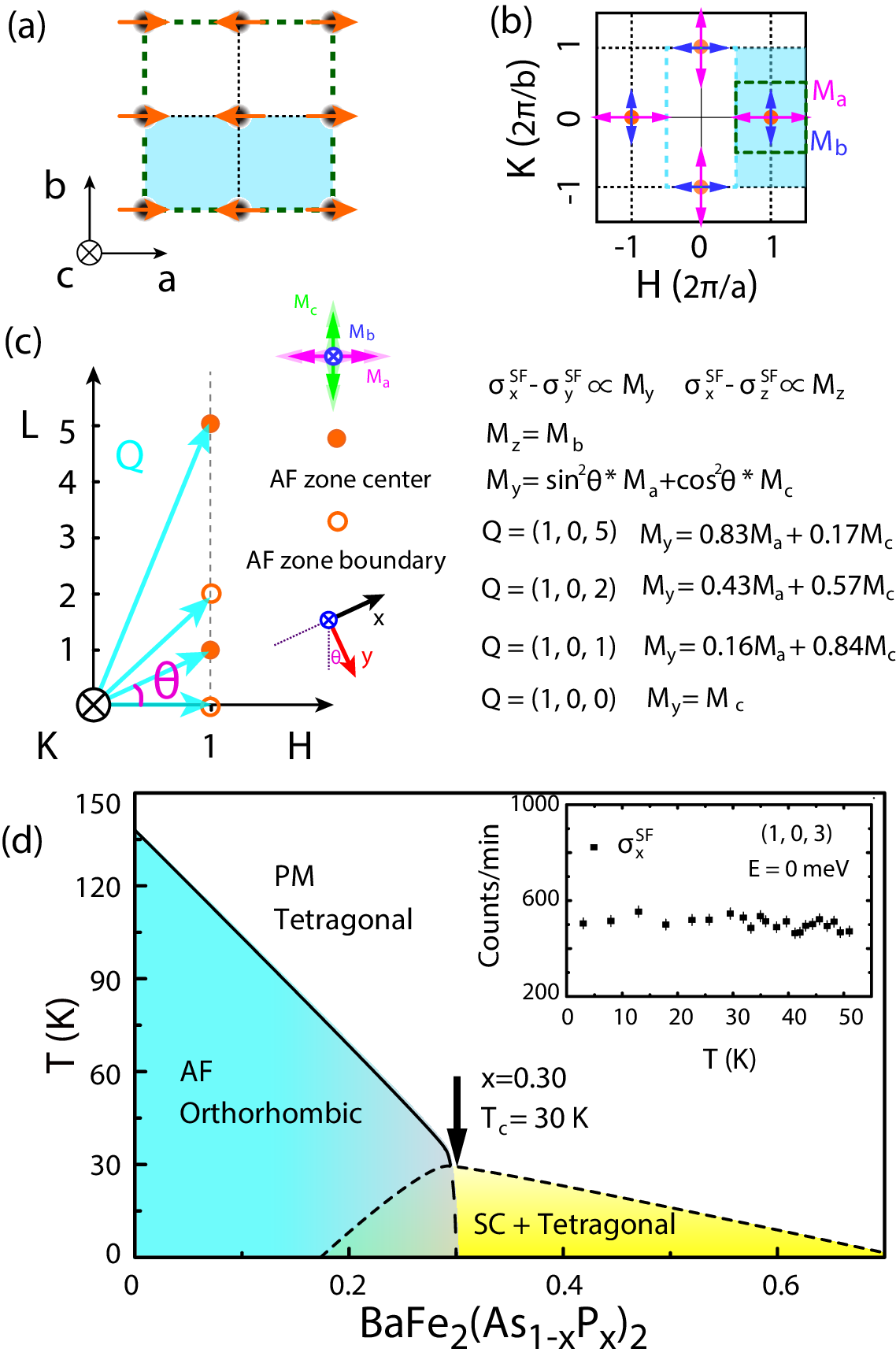}
\caption{(a) Chemical (dotted line with the orthorhombic
lattice parameters of $a$, $b$, and $c$) and magnetic (cyan area) unit cells of BaFe$_2$As$_2$ in the AF orthorhombic phase. The arrows depict the stripe magnetic structure with ordered moments aligned along the $a$-axis.  (b) The reciprocal space of BaFe$_2$As$_2$, where orange dots indicate in-plane AF ordering wave vector (${\bf Q}_{\rm AF}$). The presence of magnetic peak at $(0,\pm 1)$ is due to twinning.
Magenta and blue arrows indicate spin excitations polarized along in-plane longitudinal ($M_a$) and transverse ($M_b$) directions. (c) $[H, 0, L]$ scattering plane used in this experiment. Cyan arrows indicate the
measured ${\bf Q}$, the solid orange dots are the AF ZCs and the
orange circles are at the zone boundaries along $c$-axis.
Neutron polarization directions ($\alpha = x, y, z$) and deduced three components ($M_{a,b,c}$) of magnetic excitations
are marked by colored arrows. The equations show the relationship of $M_{a,b,c}$ and
$\sigma^{\rm NSF}_\alpha$ at different wave vectors. (d) The schematic electronic phase diagram of BaFe$_2$(As$_{1-x}$P$_x$)$_2$ \cite{DHu2015}. The black arrow marks BaFe$_2$(As$_{0.70}$P$_{0.30}$)$_2$ in phase diagram. No AF order is found at
${\bf Q}_{\rm AF}=(1,0,3)$ position as shown in the inset.}
\label{fig1}
\end{figure}

To resolve this problem, we used polarized INS to study low-energy spin excitations in optimally isovalent-doped
BaFe$_2$(As$_{0.7}$P$_{0.3}$)$_2$ ($T_c = 30$ K, Fig. 1),
where superconductivity induces a resonance with $E = 9$ meV at ${\bf Q}_{\rm AF}=(1,0,1)$ that disperses to $E = 12$ meV at ${\bf Q}=(1,0,0)$ (Fig. 2) \cite{CHLee13,HDing16}.
We chose BaFe$_2$(As$_{0.7}$P$_{0.3}$)$_2$ because the system has no AF order and structural distortion \cite{DHu2015}, and is believed to be near
a magnetic \cite{Shibauchi14} or a nematic quantum critical point \cite{HHKuo2016}. Since the AF order in BaFe$_2$(As$_{1-x}$P$_{x}$)$_2$ is gradually suppressed with increasing $x$ similar to electron- and hole-doped BaFe$_2$As$_2$ \cite{DHu2015}, one would expect low-energy spin excitations
in BaFe$_2$(As$_{0.7}$P$_{0.3}$)$_2$ to behave similarly to those of optimally doped BaFe$_{2-x}TM_x$As$_2$ and (Ba,K)Fe$_2$As$_2$, and exhibit anisotropy at temperatures well
above $T_c$ \cite{PSteffens,HQLuo2013,YSong16}.
Surprisingly, we find that spin excitations are completely isotropic in spin space above $T_c$ for energies above 2 meV. Upon entering into the superconducting state, spin excitations at ${\bf Q}_{\rm AF}=(1,0,1)$ become anisotropic, with the $a$-axis polarized resonance extending to the lowest energy ($M_a\geq 3$ meV), followed by $c$-axis ($M_c \geq 5$ meV) and $b$-axis ($M_b \geq 6$ meV) polarized modes [Figs. 3 and 4].
The resonance and spin excitation anisotropy vanish around $T_{c}$. Although superconductivity also induces a resonance at ${\bf Q}=(1,0,0)$, it is isotropic with $M_a\approx M_b\approx M_c$
[Figs. 2(d) and 2(e), 3(e)-3(h)]. These results thus indicate that the spin excitation anisotropy in BaFe$_2$(As$_{0.7}$P$_{0.3}$)$_2$ is closely related to the static AF order and superconductivity.
Assuming that SOC in iron pnictides gives rise to magnetic single-ion anisotropy that controls the ordered moment direction in the AF ordered phase \cite{NQureshi2012,CWangPRX}, the dramatic temperature dependence of the spin excitation anisotropy across $T_{c}$ with negligible modification of the lattice suggests a direct association of SOC with magnetic fluctuations inside the superconducting state \cite{AEBohmer2012}.

\begin{figure}
\includegraphics[scale=0.4]{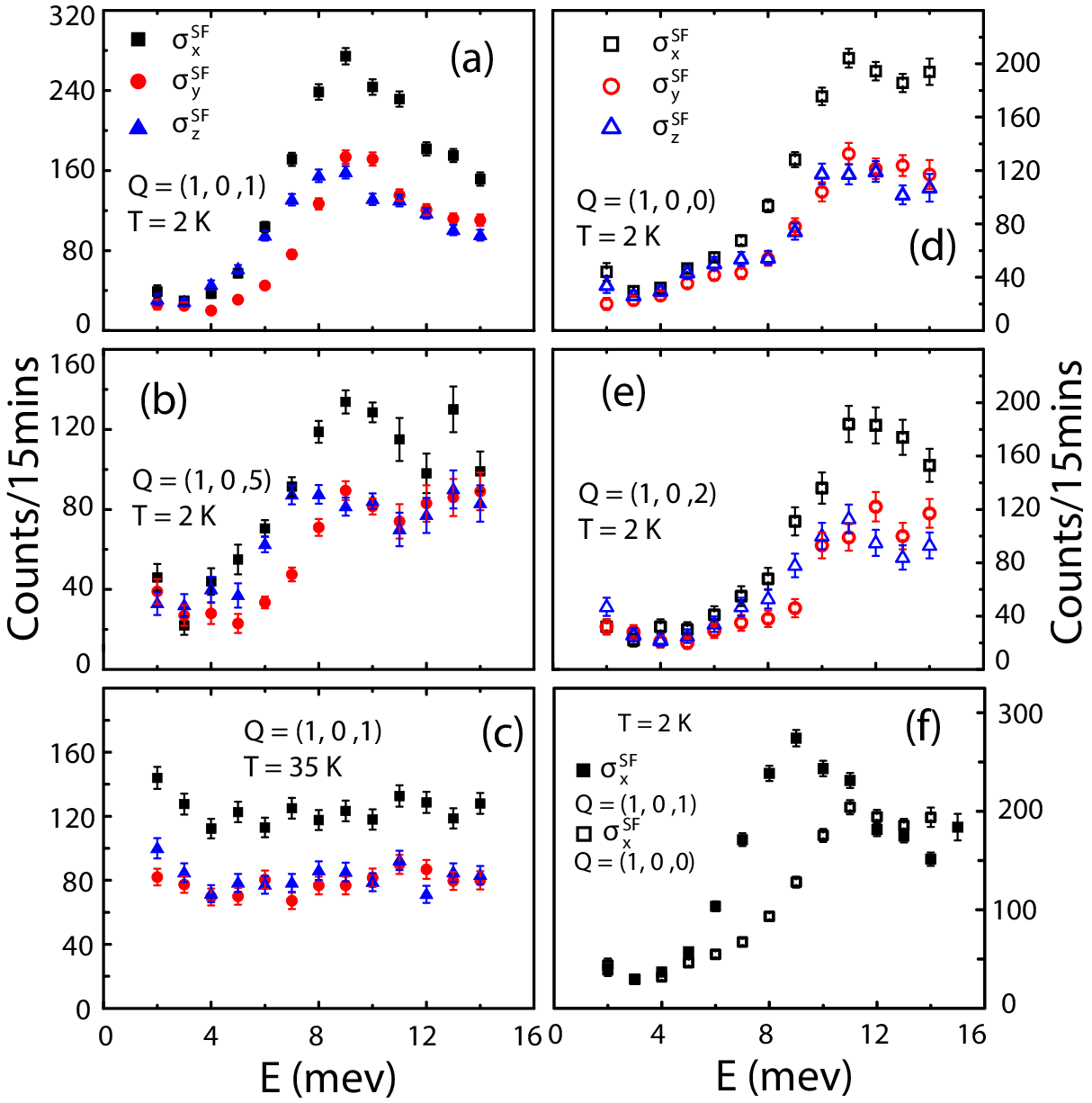}
\caption{Energy scans for the neutron SF channels under different neutron polarization directions, marked as $\sigma^{\rm SF}_{x,y,z}$, at $T = 2$ K and $\textbf{Q} = (1, 0, L)$ for (a) $L =1$, (b) $L=5$, (d) $L =0$, (e) $L =2$; and (c) at 35 K with ${\bf Q} = (1, 0, 1)$. (f) The neutron SF scattering $\sigma^{\rm SF}_{x}$ at $T=2$ K with $L = 1$, and 0. }
\label{fig2}
\end{figure}

Figure 1(b) shows the reciprocal space of magnetically ordered BaFe$_2$As$_2$ [Fig. 1(a)], where
the magnitudes of spin excitations polarized along the $a$, $b$, and $c$-axis directions at ${\bf Q}_{\rm AF}=(1,0,1)$ are marked
as $M_a$, $M_b$, and $M_c$, respectively.
Our high quality BaFe$_2$(As$_{0.7}$P$_{0.3}$)$_2$ single crystals were grown by self-flux method as described previously \cite{DHu2015}. We have co-aligned 17-g single crystals in the $[H,0,L]$
scattering plane with an in-plane mosaic $<7^\circ$. Polarized INS experiment was performed using the triple-axis spectrometer IN22 at the Institut Laue-Langevin, Grenoble, France, using a Cryopad
as described previously \cite{HQLuo2013}. In such an experiment, the incident neutron beam is polarized along the momentum
transfer ${\bf Q}$ direction ($x$) or two perpendicular directions ($y,z$) as shown in Fig. 1(c). Neutron spin-flip (SF)
and non-spin-flip (NSF) scattering cross-sections for each polarization can then be written as $\sigma^{\rm SF}_{\alpha}$ and
$\sigma^{\rm NSF}_{\alpha}$ ($\alpha = x, y, z$), respectively. The leakage between SF and NSF channels can be quantified by
the neutron spin flipping ratio $R=\sigma^{\rm NSF}_\alpha/\sigma^{\rm SF}_\alpha$ for a nuclear Bragg peak \cite{CZhang2014}. For the experiments, we find $R=16$ for all neutron polarizations. By carrying out neutron polarization analysis at ${\bf Q}_{\rm AF}=(1,0,L=1,5)$ and ${\bf Q}=(1,0,L=0,2)$
with $\sigma^{SF}_\alpha$, we can determine the magnitude of magnetic scattering $M_a$, $M_b$, and $M_c$ via
$(\sigma^{\rm SF}_x-\sigma^{\rm SF}_y)/c=M_y=M_a\sin^2\theta +M_c\cos^2\theta $ and
$(\sigma^{\rm SF}_x-\sigma^{\rm SF}_z)/c=M_z=M_b$, where $c=(R-1)/(R+1)$ and
$\theta$ is the angle between ${\bf Q}=(1,0,L)$ and $(1,0,0)$ [Fig. 1(c)] \cite{dai}. As measurements of $\sigma^{\rm SF}_\alpha$
at ${\bf Q}_{\rm AF}=(1,0,1)$ can only give $M_y$ and $M_z$, a conclusive determination of $M_a$ and $M_c$ requires data at more than
one AF ZC positions [Fig. 1(c)] \cite{HQLuo2013}.

\begin{figure}
\includegraphics[scale=0.4]{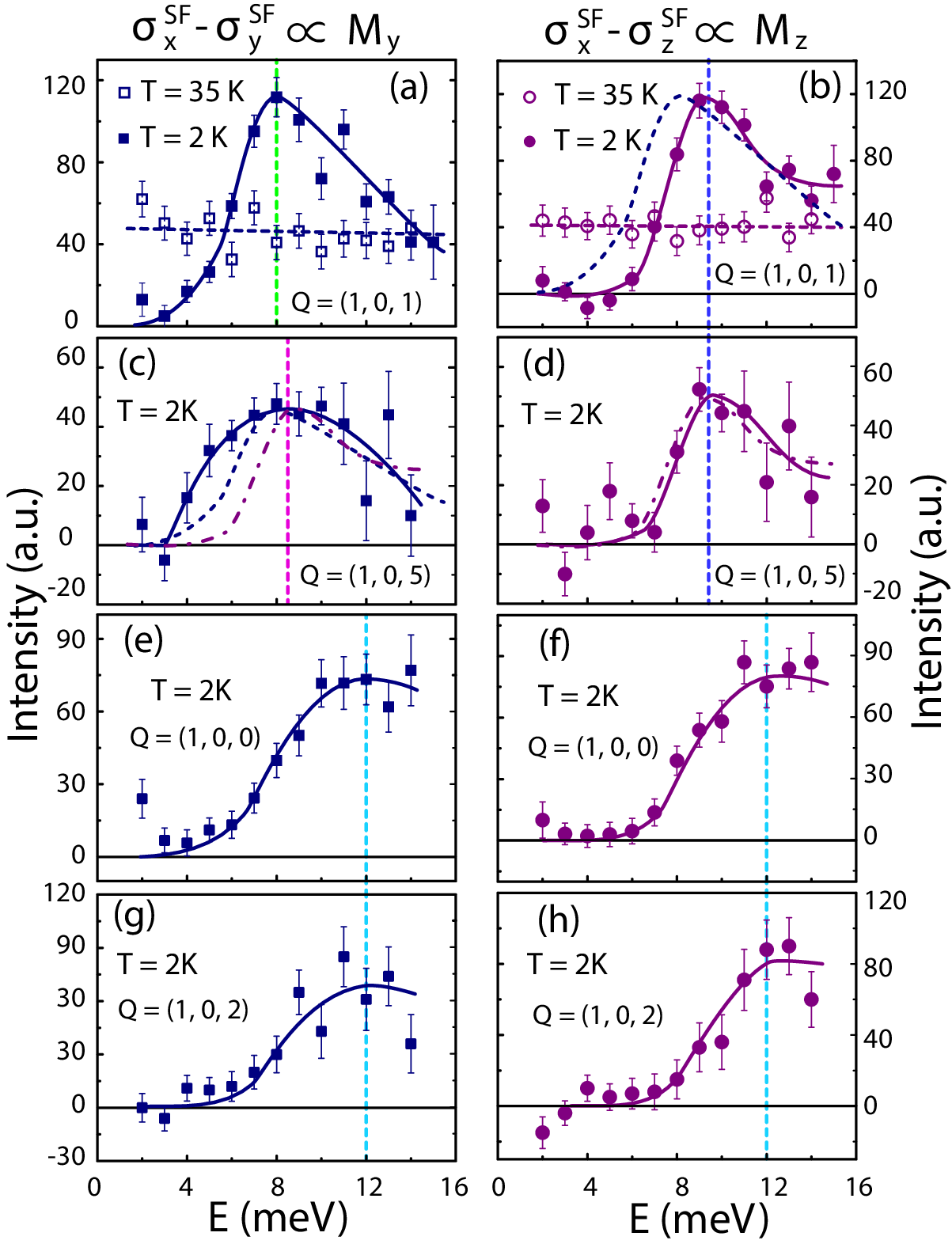}
\caption{The differences between $\sigma^{\rm SF}_{x}$ and $\sigma^{\rm SF}_{y,z}$ at different $\textbf{Q} = (1, 0, L)$ and temperatures. (a), (b) $L =1$. (c), (d) $L =5$. (e), (f) $L =0$. (g), (h) $L =2$. The solid and dashed lines are guides to the eye.
}
\label{fig3}
\end{figure}

In previous unpolarized INS experiments on optimally P-doped BaFe$_2$(As$_{1-x}$P$_{x}$)$_2$ superconductor,
the neutron spin resonance is found to be dispersive along the $c$-axis, suggesting a close connection
of the mode to the three-dimensional AF spin correlations \cite{CHLee13,HDing16}.
Figures 2(a) and 2(b) show raw data of $\sigma^{\rm SF}_{\alpha}$ in the superconducting state ($T=2$ K) at ${\bf Q}_{\rm AF}=(1,0,1)$
and ${\bf Q}_{\rm AF}=(1,0,5)$, respectively. For isotropic paramagnetic scattering with the same background in different channels, one would expect $(\sigma^{\rm SF}_{x}-BG)/2 = \sigma^{\rm SF}_{y}-BG = \sigma^{\rm SF}_{z}-BG$.
While the data show a clear resonance around 9 meV for all three neutron polarization directions ($x,y,z$),
there are clear differences between $\sigma^{\rm SF}_{y}$ and $\sigma^{\rm SF}_{z}$ below 10 meV. For energies below 3 meV,
there is no magnetic scattering due to the opening of a spin gap in response to superconductivity
($\sigma^{\rm SF}_{x} = \sigma^{\rm SF}_{y} = \sigma^{\rm SF}_{z}$).
On warming to $T = 35$ K above $T_{c}$, the scattering becomes featureless down to 2 meV with
$\sigma^{SF}_{x}/2 \approx \sigma^{SF}_{y} = \sigma^{SF}_{z}$ consistent with the scattering being isotropic [Fig. 2(c)]. These results are clearly different from those
of hole \cite{CZhang2013,NQureshi2014,YSong16} and electron-doped \cite{Lipscombe,PSteffens,HQLuo2013} BaFe$_2$As$_2$.

\begin{figure}
\includegraphics[scale=0.4]{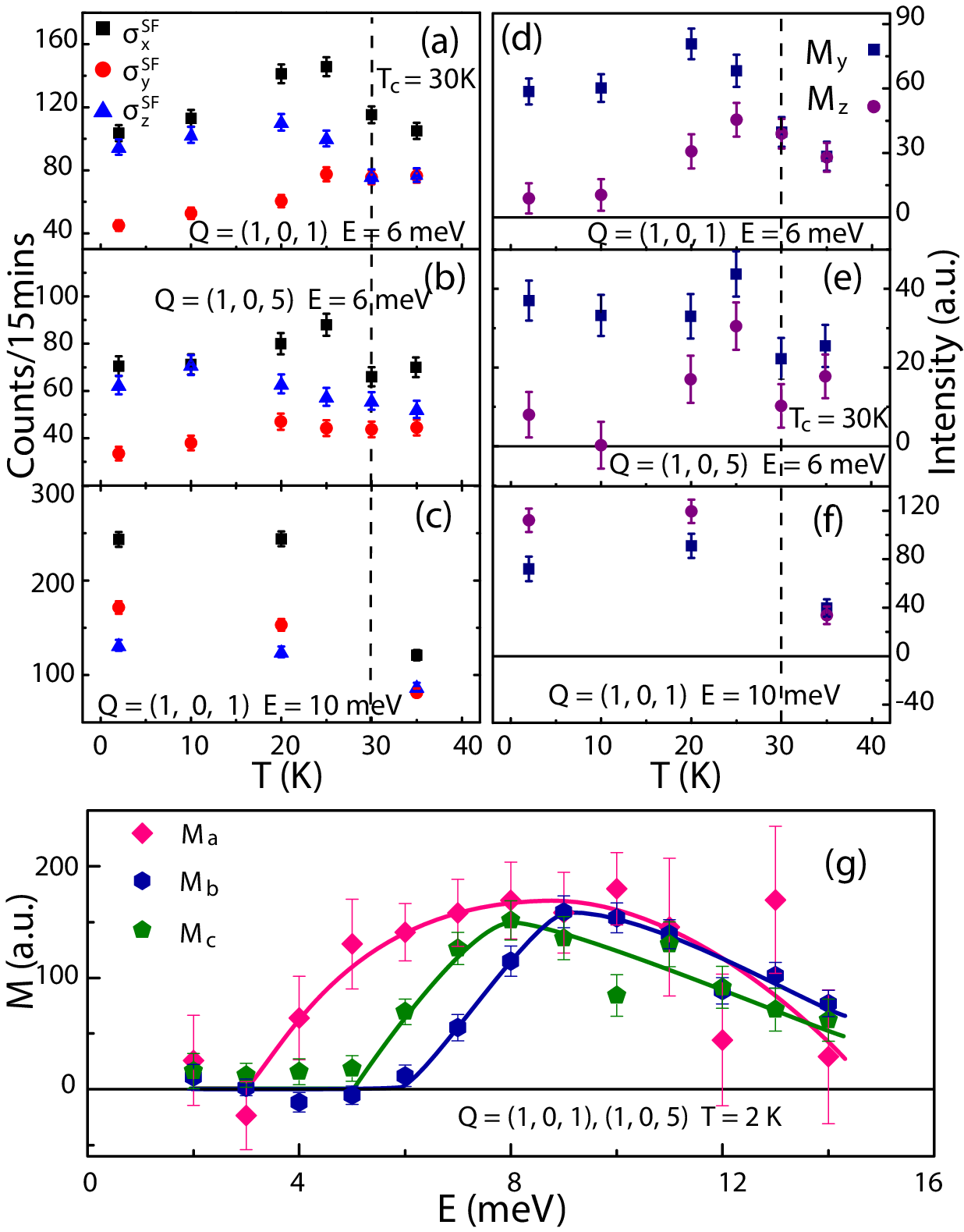}
\caption{Temperature dependence of $\sigma^{\rm SF}_{x,y,z}$ at (a) ${\bf Q}(=1,0,1)$ and  $E = 6$ meV; (b) ${\bf Q}=(1,0,5)$
and $E = 6$ meV, and (c) ${\bf Q}=(1, 0, 1)$ and $E = 10$ meV. (d)-(f) show the corresponding $M_y$ and $M_z$. (g)
The energy dependence $M_a$, $M_b$, and $M_c$ at $T=2$ K obtained using data in Figs. 2(a) and 2(b) \cite{CZhang2014}.
The vertical dashed lines mark $T_{c}$. The solid lines are guides to the eye.
}
\label{fig4}
\end{figure}

Figures 2(d) and 2(e) show $\sigma^{\rm SF}_{\alpha}$ below $T_{c}$ at the AF zone boundary
positions ${\bf Q}=(1,0,0)$ and ${\bf Q}=(1,0,2)$, respectively. While the data shows a clear resonance
around 12 meV and a spin gap below 6 meV, we find $\sigma^{\rm SF}_{x}/2 \approx \sigma^{\rm SF}_{y} \approx \sigma^{\rm SF}_{z}$
at all other energies indicative of isotropic scattering.  The dispersive nature of the resonance
is clearly seen by over-plotting $\sigma^{\rm SF}_{x}$ at ${\bf Q}_{\rm AF}=(1,0,1)$
and ${\bf Q}=(1,0,0)$ [Fig. 2(f)] \cite{HDing16}.

To gain insight into the spin excitation anisotropy from $\sigma^{SF}_{\alpha}$, we calculate the energy dependence
of $M_y$ and $M_z$ at different $L$ values, corresponding to different sensitivity to $M_a$ and $M_c$ [Fig. 1(c)].
Figures 3(a) and 3(b) show the energy dependence of $M_y$ and $M_z$ at ${\bf Q}_{AF}=(1,0,1)$ below and above $T_{c}$.
In the normal state, both $M_y$ and $M_z$ are featureless in the measured energy range. Although superconductivity
induces a resonance for both channels around 9 meV, the energy width of the resonance is narrower in $M_z$, resulting
to a smaller spin gap for $M_y$. At ${\bf Q}_{\rm AF}=(1,0,1)$, $M_y=0.16M_a+0.84M_c$ and is therefore mostly sensitive
to $M_c$. These results suggest that the $c$-axis
polarized resonance extends to lower energies than the $b$-axis polarized resonance \cite{Supplementary}.
Figures 3(c) and 3(d) plot the energy dependence of $M_y$ and $M_z$ at ${\bf Q}_{\rm AF}=(1,0,5)$ below $T_{c}$.
As $M_z=M_b$ is independent of $L$, we would expect identical energy dependence for scans at ${\bf Q}_{\rm AF}=(1,0,1)$ and
${\bf Q}_{\rm AF}=(1,0,5)$ aside from minor differences due to instrumental resolution and the Fe magnetic form factor \cite{CZhang2014}.
Inspection of Figs. 3(b) and 3(d) finds this to be indeed the case.  On the other hand, since $M_y=0.83M_a+0.17M_c$
at ${\bf Q}_{AF}=(1,0,5)$, $M_y$ in Fig. 3(c) should be mostly sensitive to the $a$-axis polarized spin excitations.
Compared with Figs. 3(a, b), Fig 3(c) reveals that the shape of $M_a$ sensitive resonance is different from $M_b$ and $M_c$ with broader peak and a spin gap of 3 meV.
Figures 3(e)-3(h) summarize
the energy dependence of $M_y$ and $M_z$ at the AF zone boundaries of
${\bf Q}=(1, 0, 0)$ and $(1, 0, 2)$ positions. Consistent with Figs. 2(d) and 2(e), these scans confirm
the isotropic nature of spin excitations at the AF zone boundary.

Given the clear experimental evidence for anisotropic spin excitations at the AF ZC below $T_{c}$ and its
absence above $T_{c}$, it would be interesting to determine the temperature dependence of the spin excitation
anisotropy. Figure 4(a) shows temperature dependence of $\sigma^{\rm SF}_{\alpha}$ at 6 meV and
${\bf Q}_{\rm AF}=(1,0,1)$. The corresponding $M_y$ and $M_z$ are shown in Fig. 4(d). Since $M_y$ is dominated by $M_c$ at this wave vector
and $M_z=M_b$, superconductivity induces magnetic anisotropy at 6 meV by enhancing $M_c$ and suppressing $M_b$.
Similarly, temperature dependence of the magnetic scattering $\sigma^{\rm SF}_{\alpha}$ at 6 meV and
${\bf Q}_{\rm AF}=(1,0,5)$ in Fig. 4(b) reveals that superconductivity
also enhances $M_a$ and suppresses $M_b$ [Fig. 4(e)]. Figure 4(c) plots temperature dependence of $\sigma^{\rm SF}_{\alpha}$
at $10$ meV and ${\bf Q}_{\rm AF}=(1,0,1)$.  From the resulting $M_y$ and $M_z$ [Fig. 4(f)], we see that superconductivity
induces a weakly anisotropic $M_c$ and $M_b$ resonance at 10 meV.

Based on measurements of spin excitation anisotropies at ${\bf Q}_{\rm AF}=(1,0,1)$ and (1, 0, 5) in Figs. 1(c) and 2, we deduce the
energy dependence of $M_a$, $M_b$, and $M_c$ in the superconducting state as shown in Fig. 4(g) \cite{HQLuo2013,CZhang2014}.
The order of resonance energies for different polarization directions in BaFe$_2$(As$_{0.7}$P$_{0.3}$)$_2$ is reminiscent of the
spin anisotropy in BaFe$_2$As$_2$, where the $a$-axis corresponds to the direction of ordered moment and is lowest in energy, followed by spin waves polarized along $c$- and $b$-axes. This suggests spin anisotropy resulting from SOC in BaFe$_2$(As$_{0.7}$P$_{0.3}$)$_2$ remains similar to BaFe$_2$As$_2$, in contrast to Ba$_{0.67}$K$_{0.33}$Fe$_2$As$_2$ in which the $c$-axis polarized resonance is the lowest in energy \cite{YSong16}. This observation is in line with the phase diagram of P- and K-doped BaFe$_2$As$_2$, whereas stripe magnetic order with ordered moment along $a$-axis is observed near optimal superconductivity in BaFe$_2$(As$_{1-x}$P$_{x}$)$_2$ \cite{DHu2015}, a double-${\bf Q}$ phase with ordered moments along $c$-axis is seen near optimal superconductivity in Ba$_{1-x}$K$_{x}$Fe$_2$As$_2$ \cite{Avci,Wasser,Allred}. Our data also suggests a progressively reduced integrated spectral weight of the resonance for $M_a$, $M_c$, and $M_b$.  Since paramagnetic scattering of BaFe$_2$(As$_{0.7}$P$_{0.3}$)$_2$ is isotropic in the normal state, our results suggest that the $a$-axis polarized resonance induced by
superconductivity gains the maximum spectrum weight, followed by $c$-axis, and $b$-axis polarized resonance modes [Fig. 4(g)], in qualitative
agreement with theoretical results that considers SOC \cite{Korshunov2013}.

In general, the symmetries of the crystallographic lattice can induce anisotropies in spin space that can determine the magnetic ordered
moment direction.  For iron pnictides that display a tetragonal-to-orthorhombic lattice distortion at $T_s$, orbital ordering
in the low-temperature orthorhombic phase is believed to play an important role in determining the $a$-axis moment direction of
the collinear AF ordered phase \cite{Applegate}. When BaFe$_2$As$_2$ is doped with P to form superconducting
BaFe$_2$(As$_{0.7}$P$_{0.3}$)$_2$, the static AF order and orthorhombic lattice distortion of the parent compounds are completely suppressed,
similar to optimally hole-doped Ba$_{0.67}$K$_{0.33}$Fe$_2$As$_2$ \cite{MWang13}.
Given that both pnictides are near optimal superconductivity with no orthorhombic lattice distortion and static AF order, orbital or nematic ordering associated with lattice distortion cannot play a direct role for the observed spin excitation anisotropy \cite{Applegate}. However, if we assume
that the resonance arises from hole and electron Fermi surface nesting \cite{hirschfeld}, the presence of SOC \cite{Borisenko} may induce hole and electron-doping asymmetry, giving rise to
a double-$\bf Q$ tetragonal AF structure with ordered moments along the $c$-axis near optimally hole-doped Ba$_{1-x}$K$_{x}$Fe$_2$As$_2$
and a simple collinear AF structure for electron-doped iron pnictides
 \cite{MChristensen}.  Since AF ordered BaFe$_2$(As$_{1-x}$P$_{x}$)$_2$ also has a simple collinear magnetic structure \cite{DHu2015},
one would expect the low-energy spin excitations
in Ba$_{0.67}$K$_{0.33}$Fe$_2$As$_2$ and BaFe$_2$(As$_{0.7}$P$_{0.3}$)$_2$ to be $c$-axis and $a$-axis polarized, respectively,
as our experiments reveal. Similarly, low-energy spin excitations in NaFe$_{0.985}$Co$_{0.015}$As contain a significant $a$-axis polarized component, reflective of
the collinear AF structure of underdoped NaFe$_{1-x}$Co$_x$As with $a$-axis being the easy-axis \cite{CZhang2014}. For comparison, recent polarized INS experiments reveal that
the resonance and the normal state spin fluctuations
in FeSe are anisotropic and have a strong $c$-axis polarized component \cite{MMa17}.

The absence of spin anisotropy in the normal state of BaFe$_2$(As$_{0.7}$P$_{0.3}$)$_2$ may be related
 to less quenched disorder compared with the K- and Ni-doped BaFe$_2$As$_2$, and is also consistent with the Curie-Weiss elastoresistance seen all the way down to $T_{c}$ \cite{HHKuo2016}. The significant spin excitation anisotropy at ${\bf Q}_{\rm AF}=(1,0,1)$ below $T_{c}$ and $E \leq 10$ meV suggests a stronger out-of-plane effective coupling relative to single-ion anisotropy energies in BaFe$_2$(As$_{0.7}$P$_{0.3}$)$_2$  as compared to Ba$_{0.67}$K$_{0.33}$Fe$_2$As$_2$ \cite{Supplementary}.
Since spin excitation anisotropy is already present in the normal state of electron- and hole-doped iron pnictide superconductors,
one cannot uniquely determine the effect of superconductivity to spin excitation anisotropy.
The absence of spin excitation anisotropy in
the normal state and its appearance below $T_c$ at the AF ZC
${\bf Q}_{\rm AF}=(1,0,1)$ in BaFe$_2$(As$_{0.7}$P$_{0.3}$)$_2$ provide the most compelling evidence
that superconductivity is coupled with SOC induced spin excitation anisotropy, and such anisotropy is associated with the nearby AF order and
can occur for iron pnictides with the negligible lattice distortion \cite{AEBohmer2012}.

The neutron scattering work at Rice is supported by the
U.S. NSF-DMR-1700081 and DMR-1436006 (P.D.). A part of the material synthesis work at Rice is
supported by the Robert A. Welch Foundation Grant No. C-1839 (P.D.).
The work at IOP is supported by the "Strategic Priority Research Program (B)" of Chinese Academy of Sciences (XDB07020300), Ministry of Science and Technology of China (2012CB821400, 2011CBA00110,2015CB921302,2016YFA0300502), National Science Foundation of China (No. 11374011, 11374346, 91221303,11421092,11574359), and the National Thousand-Young-Talents Program of China. This work was additionally supported by the Swiss State Secretariat for Education, Research and Innovation (SERI) through a CRG-grant.

\end{document}